\documentclass[twocolumn,aps,prl,showpacs,superscriptaddress,groupedaddress]{revtex4}
\usepackage{textcomp}
\usepackage{amsmath}
\usepackage{amssymb}
\usepackage{graphicx}
\usepackage{color}

\begin{document}

\title{Conditional $\pi$-Phase Shift of Single-Photon-Level Pulses at Room Temperature}

\author{Reihaneh Shahrokhshahi}
\affiliation{Department of Physics and Astronomy, Stony Brook University, New York 11794-3800, USA}
\affiliation{R. Shahrokhshahi and S. Sagona-Stophel contributed equally to this work.}
\author{Steven Sagona-Stophel}
\affiliation{Department of Physics and Astronomy, Stony Brook University, New York 11794-3800, USA}
\affiliation{R. Shahrokhshahi and S. Sagona-Stophel contributed equally to this work.}
\author{Bertus Jordaan}
\affiliation{Department of Physics and Astronomy, Stony Brook University, New York 11794-3800, USA}
\author{Mehdi Namazi}
\affiliation{Department of Physics and Astronomy, Stony Brook University, New York 11794-3800, USA}
\author{Eden Figueroa}
\affiliation{Department of Physics and Astronomy, Stony Brook University, New York 11794-3800, USA}

\begin{abstract}
The development of useful photon-photon interactions can trigger numerous breakthroughs in quantum information science, however this has remained a considerable challenge spanning several decades. Here we demonstrate the first room-temperature implementation of large phase shifts ($\approx\pi$) on a single-photon level probe pulse (1.5us) triggered by a simultaneously-propagating few-photon-level signal field. This process is mediated by $Rb^{87}$ vapor in a double-$\Lambda$ atomic configuration. We use homodyne tomography to obtain the quadrature statistics of the phase-shifted quantum fields and perform maximum-likelihood estimation to reconstruct their quantum state in the Fock state basis. For the probe field, we have observed input-output fidelities higher than 90$\%$ for phase-shifted output states, and high overlap (over 90\%) with a theoretically perfect coherent state. Our noise-free, four-wave-mixing-mediated photon-photon interface is a key milestone towards developing quantum logic and nondemolition photon detection using schemes such as coherent photon conversion.
\end{abstract}

\maketitle

\section{Introduction}
Photons are famously useful for transmitting \cite{PhysRevLett_115_040502, Liao2017} and storing \cite{Wang2017, PhysRevLett.118.220501} quantum information due to their typically weak interaction with the environment. Because of this, it has been thought that photon-based quantum computers would be one of the best architectures if an efficient photon-photon gate could be developed \cite{PhysRevA.52.3489}. Additionally, there are a number of unique applications for photonic quantum information processing, including utilization of qRAM \cite{Biamonte2017} using stored qubits \cite{Lvovsky2009}, higher connectivity than current grid-based systems \cite{Lechnere_2015} using nonlocal connections between flying qubits, and continuous-variable quantum machine learning \cite{1806.06871}.
\footnote{Corresponding authors: stevensagona@gmail.com, eden.figueroa@stonybrook.edu}

There are a number of challenges towards experimentally realizing a photon-photon phase gate: i) A system first must exhibit large enough cross-phase modulation (XPM) per photon such that a single photon causes a $\pi$ phase shift on a second photon state. ii) The quantum state of the phase-shifted light must not be distorted or impure. iii) Each possibility in the ``truth table" of input combinations must be have high-fidelity operation. In particular the state should remain the same when the phase-shifting trigger photon is not present:
\begin{align*}
    (|0\rangle + |1\rangle)\otimes|1\rangle &\Rightarrow (|0\rangle - |1\rangle)\otimes|1\rangle \\
    (|0\rangle + |1\rangle)\otimes|0\rangle &\Rightarrow (|0\rangle + |1\rangle)\otimes|0\rangle
\end{align*}

Recently, cavity-based and Rydberg-based systems have demonstrated large XPM at the single-photon level \cite{Tiarks2016,Beck2016,Feizpour2015}, and have begun investigating operation fidelities of a photon-photon gate \cite{Tiarks2019, Rempe16}.
However, these new schemes all either need cavities or require the light to be stored in an atomic medium, which dramatically lowers the gate's efficiency.
For example, the most advanced photonic processing system currently has an efficiency ranging from 0.5\% to 8\%, with an entangling gate fidelity of $ 63.7\% \pm 4.5\% $  \cite{Tiarks2019}. It is clear that a lot of improvement needs to be made before these systems are ready for large problems requiring fault tolerance \cite{PhysRevLett.117.060505}. Additionally, improvements may need new physical architectures, as it is not clear if current gate implementations have fundamental upper limits on fidelities similar to the constraints present in Kerr systems \cite{PhysRevA.73.062305, PhysRevA.81.043823}.

Recent developments have shown that, unlike previously thought \cite{PhysRevA.73.062305, PhysRevA.81.043823}, there are indeed cavity-free nonlinear systems that are capable of large phase shifts at high-fidelity \cite{PhysRevLett.116.023601}, and they can perform gate operations even in a full frequency-mode framework \cite{PhysRevA.97.032314}. These systems use a paradigm called ``coherent photon conversion (CPC) \cite{Langford2011}," in which photons are converted by a four-wave mixing (FWM) process. It has been shown that universal quantum computation is possible with only a combination of these processes and linear optics \cite{PhysRevLett.120.160502}.

To match the requirements of a quantum photonic gate using CPC, the following experimental benchmarks must be achieved: i) Four-wave-mixing processes can occur for single photon inputs. ii) The fidelity of the quantum output state in the Fock state basis is preserved. iii) The two photon inputs must be efficiently and coherently converted into a third (initially vacuum) field and then be coherently converted back, thereby creating a ``Rabi oscillation" between the $|1, 1, 0\rangle$ and $|0,0,1\rangle$ states \cite{Langford2011}.

While current work towards CPC uses nonlinear waveguides and is limited by the efficiency of the quantum process \cite{PhysRevA.90.043808}, atomic systems can create FWM \cite{Liu2017} at near unitary efficiencies, and are an excellent candidate for highly efficient CPC. Additionally it has been shown that with these same FWM systems, it is possible to achieve large XPM at the low light level without requiring cavities, storage or Rydberg levels through a double-$\Lambda$ system \cite{Yu2016}. Unlike Kerr-nonlinearity-based schemes, which have increasing phase shifts per photon when the signal field coupling is increased, double-$\Lambda$ systems can have phase shifts that are independent of signal power, as long as the probe and signal are scaled simultaneously.

In this experiment we demonstrate the first room-temperature implementation of a double-$\Lambda$ system in which simultaneously propagating pulses of single-photon-level light create $\pi$ phase shifts.
We perform a quantum characterization of these phase-shifted output states using quantum state tomography of the quadrature statistics.
We show that four-wave mixing processes at room-temperature still produce well-behaved quantum states, observing high fidelities even for single-photon level light phase shifted by $\pi$. Finally, we evaluate our system in the context of the first two of the mentioned benchmarks required for implementation of the CPC protocol.

\section{Theoretical Background}
A key feature of our double-$\Lambda$ system is that the output phase-shift is phase-sensitive on the input phases of multiple fields. First, we will highlight this sensitivity theoretically and identify a key parameter $\Delta \phi_{FWM}$. Second, after outlining our experimental atomic system, we will discuss how a unique pulse-sequence protocol can be used to simultaneously extract both phase-shifts and quadrature statistics after interaction with the medium. Third, we verify these experimental estimates of phase-shifts are both precise and well aligned with our theoretic atomic model. Fourth, we will use these estimates to collect phase-shift-binned quadratures statistics for quantum state estimation. To understand the quantum behavior of the DL process at the single-photon-level, using the results of our quantum state tomography, we calculate characterization values including purity, fidelity, and coherent-state overlap.

\begin{figure*}
\includegraphics[width=1.0\linewidth]{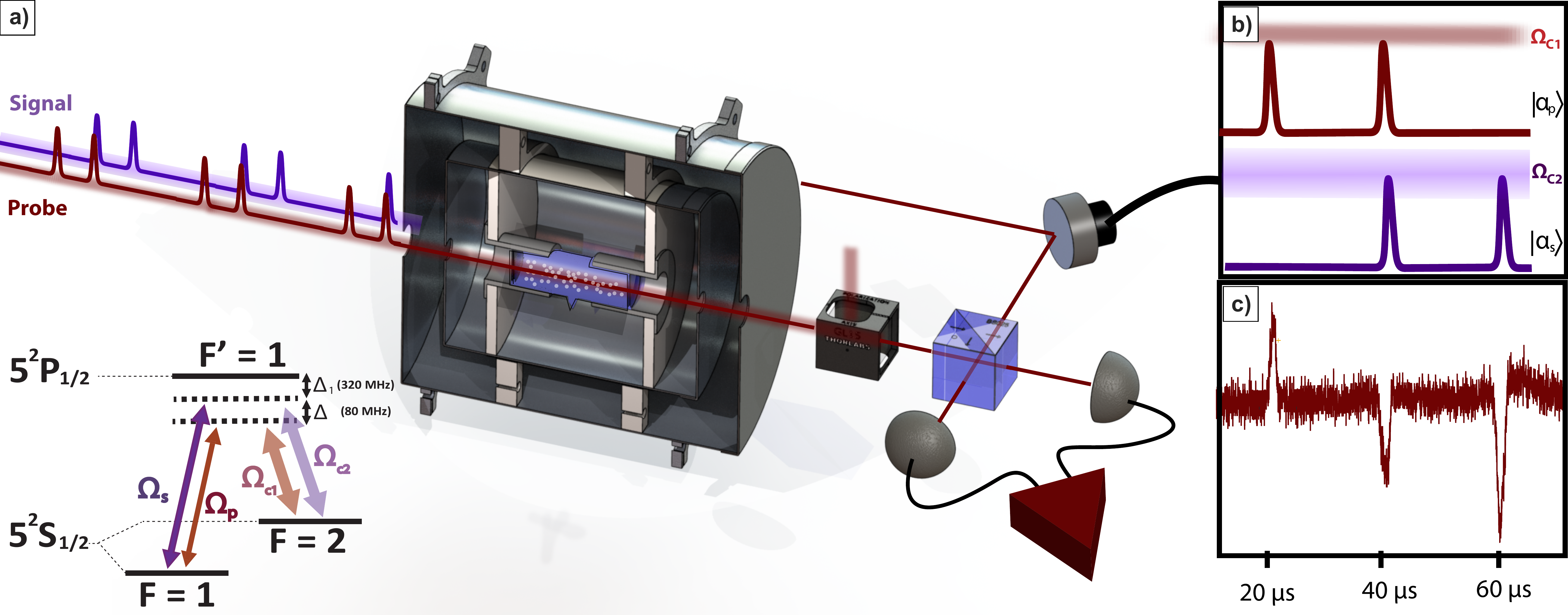}
\caption{\textbf{a)} Experimental setup. A sequence of probe ($\Omega_p$) and signal pulses ($\Omega_s$) are sent through a room-temperature $^{87}Rb$ atomic ensemble. The probe output is then measured on a balanced homodyne detector and the associated quadrature and phase values are analyzed. \textbf{b)} Diagram of our sequence of input pulses. First, we send a pulsed probe field  ($\Omega_{p}$), with two CW control fields ($\Omega_{c1}, \Omega_{c2})$ creating EIT.  Secondly, we add a pulsed signal $\Omega_{s}$ field (in addition to the probe $\Omega_{p}$ pulse), creating a double-$\Lambda$ system. Thirdly, we switch off the probe pulse (leaving the signal pulse on), creating four-wave mixing FWM in the frequency of the probe.  \textbf{c)} Voltage values of the balanced homodyne detector, representing quadrature measurements of the pulse sequence. Across a scan of the phase of the local oscillator of the homodyne, phase information can be extracted from these pulses (as discussed in Figure 2).}
\end{figure*}

\subsection{Double-$\Lambda$ System}

We use a double-$\Lambda$ system similar to the closed-loop scheme presented in \cite{Yu2016}. While a fully quantized model of similar systems have been constructed in \cite{Artoni2015}, we introduce a simpler semi-classical model based on a three-level EIT system to understand how this atom-light interaction creates a phase shift aided by four-wave mixing.

As seen in Figure 1, each individual $\Lambda$ system has a weak field coupling states $|1\rangle$ and $|3\rangle$ (named ``probe'' and ``signal'') and a strong control field (labeled ``c1'' and ``c2'') coupling states $|1\rangle$ and $|2\rangle$, with detunings ($\Delta_{p}=\Delta_{c1}=\Delta_1$) and ($\Delta_{s}=\Delta_{c2}=\Delta_2$) such that $\Delta_2-\Delta_1=\Delta$. Since our detunings are much smaller than our laser frequencies ($\Delta \ll \omega$), we can make an adiabatic approximation for the contribution of the other fields. Identifying that our additional time-dependencies are slowly-varying relative to the field's frequency, as similarly treated in standard treatment of EIT \cite{RevModPhys.77.633}, the solutions can be found via the Optical-Bloch equations:
\begin{align}
\rho_{13} &=  \frac{i \Omega_i}{2i \Delta_2+\Gamma} - \frac{\Omega_i|\Omega_j|^2}{\gamma(2i\Delta_2+\Gamma)^2}\\
\rho_{12} &= \frac{i \Omega_i \Omega_j^*}{\gamma(2i \Delta_2+\Gamma)^2} \label{eq:coherence}
\end{align}

Where $\Omega_j = E_{c1} e^{i \phi_{c1}}+E_{c2} e^{i \Delta_1 t + \phi_{c2}}$ and $\Omega_i = E_p e^{i \phi_{p1}}+E_s e^{i \Delta_1 t + \phi_{s2}}$ are effective Rabi frequencies containing the time dynamics of both fields. Explicitly plugging in these time dependent relations, and solving for frequency terms that match the probe and signal, we obtain the following relations for the output polarizability:

\begin{widetext}
\begin{align}
P(\omega_s) &= e^{i\phi_s}\left(\left(\frac{i |\wp_{13}|}{2i \Delta_2+\Gamma} + \frac{|\wp_{13}| |\wp_{23}|^2 (|E_{c1}|^2+ |E_{c2}|^2)}{\gamma (2i \Delta + \Gamma)^2}\right)|E_s| +\left(\frac{|\wp_{13}| |\wp_{23}|^2 |E_{c1}| |E_{c2}| }{\gamma (2i \Delta + \Gamma)^2}\right) |E_p| e^{i\Delta \phi^{s}_{FWM}}\right)\\
P(\omega_p) &= e^{i\phi_p}\left( \underbrace{ \left(\frac{i |\wp_{13}|}{2i \Delta_2+\Gamma} + \frac{|\wp_{13}| |\wp_{23}|^2 (|E_{c1}|^2+ |E_{c2}|^2)}{\gamma (2i \Delta + \Gamma)^2}\right)|E_p|}_{\text{Probe-only Contribution}}  +  \underbrace{\frac{|\wp_{13}| |\wp_{23}|^2 |E_{c1}| |E_{c2}| }{\gamma (2i \Delta + \Gamma)^2} |E_s| e^{i\Delta \phi_{FWM}}}_{\text{Four-wave-mixing Contribution}}\right)
\end{align}
\end{widetext}

Where P($\omega_s$) and P($\omega_p$), obtained from $P_{out} = Tr[\rho \wp]$, are the terms proportional to $e^{i\omega_s t}$ and $e^{i\omega_p t}$ respectively. Additionally, $\wp$ is the dipole operator, $\Gamma$ is the decay rate from the excited state, and $\gamma$ is the decoherence of $\rho_{12}$. We identify a critically important parameter for our experiment: $\Delta \phi_{FWM} = \phi_{s} - \phi_{p} + \phi_{c2} - \phi_{c1}$, which represents the relative phase between the generated four-wave-mixing and the probe.  While the first term of equation 4 represents the probe light experiencing an effect similar to electromagnetically-induced transparency (EIT) due to the contribution of the strong control fields, the third term in equation 4 represents a phase-sensitive four-wave-mixing contribution, in which the two control fields and the signal create light in the frequency of the probe. Thus, the output polarizability can experience constructive or destructive interference depending on the value of $\Delta \phi_{FWM}$.

These polarizabilities contribute ``effectively linear'' driving terms to the slowly-varying Maxwell-Schrödinger equations and, similar to other work, can be analytically integrated to obtain the electric field for the probe and the signal \cite{Yu2016, Artoni2015}.

\begin{align}
\frac{\partial E_p}{\partial z} + \frac{1}{c} \frac{\partial E_p}{\partial t} = i \frac{\alpha_p \gamma_{31}}{2 L \mu_{13}} P(\omega_p) \\
\frac{\partial E_s}{\partial z} + \frac{1}{c} \frac{\partial E_s}{\partial t} = i \frac{\alpha_s \gamma_{31}}{2 L \mu_{13}} P(\omega_s)
\end{align}

This phase-sensitivity of the FWM is an extremely critical feature of the double-$\Lambda$ system because the output light's amplitude $E_p^{DL}(\Delta \phi_{FWM})$  and phase shift (written as $\Delta \phi_{DL}$) are both functions of $\Delta \phi_{FWM}$:

\begin{align}
\Delta \phi_{DL}^{\omega_p} &= \arctan{\left(\frac{
\Im [E_p^{DL}(\Delta \phi_{FWM})]}{\Re [E_p^{DL}(\Delta \phi_{FWM})]}\right)}  \label{eq:phase}
\end{align}

Therefore, for each $\Delta \phi_{FWM}$ there is an associated double-$\Lambda$ phase-shift $\Delta \phi_{DL}$ and a particular value of  $|E_p^{DL}|^2$, which will be proportional to mean-photon number for quantum light (discussed further in the next section and illustrated in Figure 2c and 3a).

\begin{figure*}
\centerline{\includegraphics[width=1.95\columnwidth]{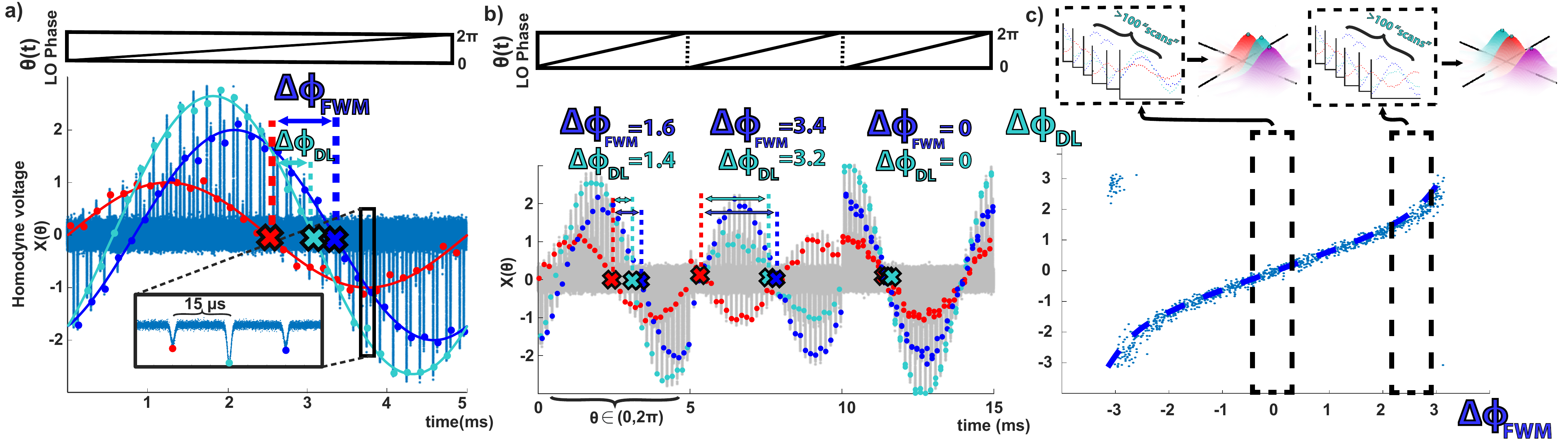}}
\caption{\textbf{Illustration of data extraction workflow for phase estimation and quantum state estimation.} \textbf{a)} For each ``scan," corresponding to a 5ms long sweep of the local oscillator phase of the homodyne, our three-pulse sequence, shown in the inset (and in Figure 1b), repeats tens of times. Peak voltage values obtained within each of these scans provides a set of quadrature values: $\{X(\theta_i) | \theta_i\in(0,2 \pi)\}$. These voltage values represent a set of quadrature measurables for the cases: probe only $\{X(\theta)\}^{\phi_p}_{p}$, double-$\Lambda$ $\{X(\theta)\}^{\phi_{DL}}_{DL}$, and four-wave mixing $\{X(\theta)\}^{\phi_{FWM}}_{FWM}$, shown as red, teal, and blue points respectively. Across each 5ms homodyne ``scan," \textit{phase-changes} in that time interval can be estimated. The output phase shift of the double-$\Lambda$ system, called $\Delta\phi_{DL}$, (relative to the probe-only case) is labeled in teal. \textbf{b)} Quadrature values are taken continuously over multiple seconds. The local oscillator ``scans" hundreds of times, generating hundreds of sets of quadrature values $\{X(\theta)\}$. Each set has a fixed associated phase-shifts ($\Delta\phi_{FWM}, \Delta\phi_{DL}$), and a range of local oscillator phases ranging from $(0, 2 \pi)$.  \textbf{c)} The relationship between DL phase shift, $\Delta\phi_{DL}$, and FWM phase shift, $\Delta\phi_{FWM}$, for each quadrature set are plotted as blue points, while the theoretical relationship is plotted as a blue-dashed line. Furthermore, the binning workflow is illustrated in the figure: Hundreds of quadrature sets associated with particular values of $\Delta \phi_{FWM}$ are binned and collected for quantum state estimation. For each phase-shift bin, the quantum states of these sets $(\{X_{P}\}, \{X_{DL}\}, \{X_{FWM}\})$ are reconstructed using maximum likelihood estimation and are illustrated as wigner functions for the EIT, FWM, and DL cases (illustrated in further detail in Figure 4).
}
\end{figure*}

\section{Experimental Methods}
\subsection{Experimental Setup}
As illustrated in Figure 1, we use a double-$\Lambda$ atomic system in a room-temperature $Rb^{87}$ atomic ensemble, in which both individual $\Lambda$ subsystems exhibit EIT. The output light is sent to a homodyne detector to extract out the quantum state. In addition to using standard techniques to extract out phase-information and quadrature statistics \cite{lvovtecniques, lvovsky2009rev}, we incorporate a unique pulse-sequence protocol to extract out quadrature statistics of the double-$\Lambda$ system corresponding to particular values of $ \Delta \phi_{FWM}$.

As shown in Equation \ref{eq:phase}, because our double-$\Lambda$ system's output light's amplitude $E_p^{DL}$ and phase shift $\Delta \phi_{DL}^{\omega_p}$ are functions of the parameter $\Delta \phi_{FWM}$, in order to accurately recover homodyne statistics describing a particular quantum state of the phase-shifted double-$\Lambda$ system, we must either have extremely precise phase-control of all of our input field's phases or have a method of accurately estimating the phases for $\Delta \phi_{FWM}$ and $\Delta \phi_{DL}$ simultaneously. We achieve the ladder by implementing a closely-separated, three pulse sequence with a fast local oscillator scan (as illustrated in Figure 2). With this implementation, we can let our phases fluctuate randomly and collect quadrature statistics of our output double-$\Lambda$ system and bin them by the phase-shift that uniquely describes their state ($\Delta \phi_{FWM}$).
Additionally, our phase-extraction protocol also allows for additional quadrature data to be collected describing the quantum states of the light generated by the FWM (denoted $\rho_{FWM}$) and the probe light experiencing EIT-like behavior without the inclusion of signal field (denoted $\rho_{EIT}$).

\begin{figure*}
\centerline{\includegraphics[width=2\columnwidth]{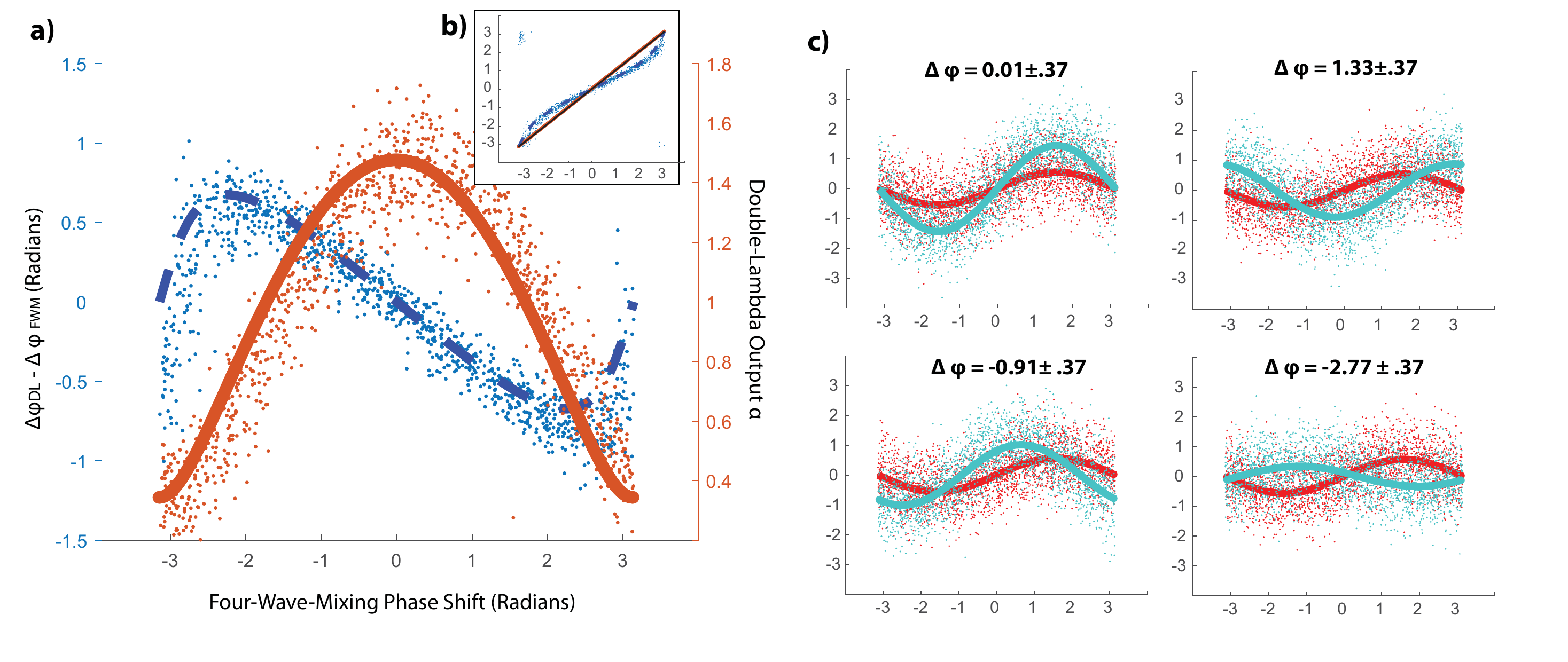}}
\caption{\textbf{In a,b)}: \textbf{Comparison between the phase and amplitude of each sinusoidal fit of individual homodyne scans and theoretical values for single-photon-level data}.  a) The blue dots represents individual values of $\Delta \phi_{DL} - \Delta\phi_{FWM}$ vs $\Delta\phi_{FWM}$ for each individual homodyne ``scan'' (as discussed in Figure 2). The orange dots represent the fitted amplitudes of each homodyne shot for the DL case. These are compared to their theoretical values (solid orange and dashed blue lines) obtained from equations (14) and (15). No free parameters are used in the theoretical fits. In \textbf{b)} the blue points represent the phase-shift, $\Delta\phi_{DL}$ vs $\Delta\phi_{FWM}$. Part \textbf{c)}, to reconstruct the quantum state, the quadrature statistics obtained from the outputs must be sorted by their associated $\Delta\phi_{FWM}$ phase-shifts. Using the phase information illustrated in a and b, we can collect quadrature statistics associated with different binned phases (as discussed in Figure 2). Extracted single-photon-level quadrature data illustrated for the double-$\Lambda$ (teal) and EIT reference (red) - for different sets of post-selected phase shifts (for bins $\Delta\phi_{FWM} \in .01, 1.33, -.91,$ and -2.77 $\pm .37$ radians). Quadrature values are compared to their averaged fitted values (plotted as a solid line in red and teal).
}
\end{figure*}

\subsection{Atomic Scheme}
 In our experiment, two hyperfine ground states are coupled to an excited state by weak probe fields called probe and signal ($\Omega_p$ and $\Omega_s$) and two strong control fields ($\Omega_{c1}$ and $\Omega_{c2}$), as shown in Figure 1.  We utilize two external cavity diode lasers phase-locked at 6.8 GHz to correspond to the $^{87}$Rb ground state splitting. For the first $\Lambda$ system, the probe and control field are 400 MHz red detuned from the $|5S_{1/2}, F = 1\rangle$ $\Rightarrow$ $|5P_{1/2}, F = 1 \rangle $  and the $|5S_{1/2}, F = 2\rangle$ $\Rightarrow$ $|5P_{1/2}, F = 1 \rangle $ transition respectively. The signal and second control field for the second $\Lambda$ system are 80 MHz blue detuned from the probe and first control field respectively. Both probe and signal fields are fixed to have the same linear polarization, while the control fields will be set to have the same orthogonal polarization to allow for convenient separation of the control fields after atomic interaction. The intensity of the probe and signal fields are temporally modulated by means of acousto-optical modulators (AOM).
 For our medium we use an 5 cm long glass cell with anti-reflection coated windows containing isotopically pure $^{87}$Rb with 10 Torr of Ne buffer gas kept at a temperature of 336 K. In this medium, 1.5 us long single-photon pulses can be transmitted through the EIT medium with 80$\%$ efficiency in the presence of the first control field $\Omega_{c1}=45 mW$. Additionally, with the addition of the second control field $\Omega_{c2}=5 mW$, this drops to 50$\%$.

\subsection{Homodyne Detection of Pulse Sequence}
As illustrated in Figure 1 b, we send a pulse sequence consisting of three cases: probe pulse only, probe and signal pulse on (creating the double-$\Lambda$ system), and signal only (generating FWM). Each pulse is $1.5 \mu s$ long, each case is separated with a 20$\mu s$ interpulse delay, and this is repeated every $60 \mu s$. Control fields 1 and 2  are on during the whole cycle.

 We use a standard balanced homodyne detector to obtain accurate phase and quadrature information to estimate the quantum state of the phase shifted light \cite{lvovsky2009rev, lvovtecniques}.
A homodyne detector can obtain phase information of cw light through interference with a strong coherent source, producing an output voltage of the form:
\begin{align}
V_{homo} &= E_p E_{LO} \cos(\phi_{LO} - \phi_p ) \\
&= E_p E_{LO}  \cos(\omega_{pz} t - \phi_p)        \label{eq:2}
\end{align}

where $E_p$ ($E_{LO}$) and $\phi_{p}$ ($\phi_{LO}$) are the amplitude and phase respectively of the probe (local oscillator). The phase of the probe, $\phi_p$, relative to the local oscillator can be obtained by scanning the local oscillator phase (with a frequency of $\omega_{pz}$) and fitting the resultant voltage $ V(t) \propto \cos(\omega_{pz} t - \phi_p) $ to a cosine curve to extract the offset, $\phi_p$.

Using our pulse sequence, we can extract out this sinusoidal behavior by extracting the voltage values associated with the peaks of the pulses. The peaks of each case in the pulse sequence forms its own sinusoidal function per scan of the local oscillator, and can be seen as the red, teal and blue fits in Figure 2a for $\phi^p$, $\phi^p_{DL}$, and $\phi^p_{FWM}$.

As a result, we can obtain the phase-shifts $\Delta \phi_{DL}$ and $\Delta \phi_{FWM}$:

\begin{align}
 \Delta \phi_{DL} &= (\phi^p_{DL}-\phi_{LO}) - (\phi_{p} - \phi_{LO}) = \phi^p_{DL} - \phi_{p} \\
 \Delta \phi_{FWM} &= (\phi^p_{FWM}-\phi_{LO}) - (\phi_{p} - \phi_{LO}) = \phi^p_{FWM} - \phi_{p}
\end{align}

By alternating between the three cases and quickly scanning the local oscillator phase, we can obtain a phase estimate faster than the timescale of the phase-fluctations of our experiment (typically due to slight optical path length changes due to air fluctations). Using a Spectrum M2i.2031 digitizer card with 2 GSamples of onboard memory, we store continuous homodyne data in bursts of 1.3 seconds, with an internal sampling clock of 100MHz. We acquire these large bursts of data, split the data into individual 5ms ``shots'' (corresponding to the time to make a complete $2\pi$ cycle of the local oscillator phase), and obtain the phase-shift per shot from each fit (illustrated in Figure 2b). Because the frequency of our local oscillator scan is 200 Hz (a frequency faster than typical phase fluctuations in our experiment), each fit provides an accurate shot-by-shot estimation of the phases of the light pulses within the 5ms time window.

Finally, we can extract the homodyne quadrature statistics for any of the outputs of the three cases in the sequence. For single-photon level light, the voltage output associated with the peak of the pulses (in each sequence) of the homodyne is quantum mechanically a measurement of the generalized quadrature:
\begin{equation*}
X_{\theta}= ae^{i \theta}+ a^{\dagger}e^{i \theta}
\end{equation*}

\begin{figure*}
\centerline{\includegraphics[width=2\columnwidth]{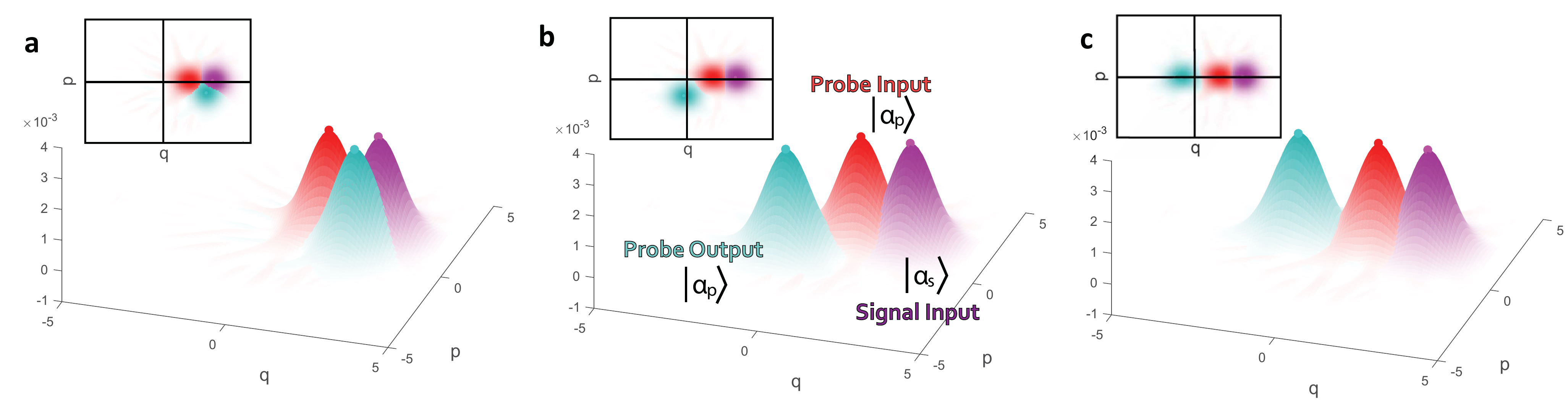}}
\caption{Reconstructed density matrices represented as Wigner functions are plotted, illustrating the relationship between inputs and outputs for different post-selected phase-shifts. The red and purple plots are the input states of the probe and signal. The teal plot is the phase-shifted probe output for \textbf{a)} 0.3 radians phase shift, \textbf{b)} $\pi/2$ phase shift, and \textbf{c)} $\pi$ phase shift.}
\end{figure*}

These quadrature statistics can be collected and binned by the assocated phase shift $\Delta \phi_{FWM}$, where the data across multiple local oscillator scans can be evaluated using Maximum-likelihood estimation to uncover their quantum states in the Fock state basis \cite{lvovsky2009rev}. This workflow is illustrated in Figure 2c and the resultant binned data is shown in Figure 3c.

\section{Results}
\subsection{Analysis of Input-phase vs. Phase-shift}

As our local oscillator sweeps through its phase, the output voltage associated with the peaks of the output pulses in the sequence vary sinusoidally (illustrated in Figure 2a and 2b). To compare our experimental results with our semi-classical theory, we can predict this behavior as a linear inteference of the ``probe-only'' contribution (the first term in equation 3) and ``FWM'' contribution (second term):

\begin{align}
E_{DL}(t) &= E_{E} \sin(w_{pz}t) + E_{F} \sin(w_{pz}t+\Delta\phi_{FWM})\\
       &= A(E_{F}, E_E, \Delta\phi_{FWM}) \sin(w_{pz}t+\Delta\phi_{FWM})
\end{align}

Where $E_E$ and $E_F$ correspond to amplitudes obtained from sinusoidal fits of homodyne shots extracted from the first and third pulse in the sequence respectively. We solve for the phase $\Delta \phi_{DL}$ and amplitude $A(E_{F}, E_E, \Delta\phi_{FWM})$, obtaining:

\begin{align}
\Delta \phi_{DL} = \arctan\left(\frac{1+\frac{E_F}{E_E}\cos(\Delta\phi_{FWM}) }{\frac{E_F}{E_E} \sin(\Delta\phi_{FWM})}   \right) \\
A = \sqrt{\left(\frac{E_{E}^2+E_{F}^2}{E_{E}^2}+2\frac{E_{F}}{E_{E}}\cos(\Delta\phi_{FWM})\right)}
\end{align}

$E_E$ and $E_F$ can be obtained accurately by averaging fits of shots of homodyne data from pulses 1 and 3 in our pulse sequence. One of each individual shot to be averaged is illustrated as red and blue sinusoidal fits in Figure 2a. Thus our simple model matches the experimental data without any free parameters at the single-photon level. The theoretical amplitude vs $\Delta\phi_{FWM}$ in equation 15 is visualized as the solid orange line in Figure 2a, while the dashed blue line in Figure 2b represents amplitude A vs $\Delta\phi_{FWM}$ from equation 14. Additionally, the small degree of uncertainty in our experimental estimation of our phases ($\Delta \phi_{FWM}$ and $\Delta \phi_{DL}$) implies we can accurately associate quadrature statistics with a particular phase-shifted ouput state, thereby allowing accurate quantum state estimation through binning.

\subsection{Quantum State Reconstruction for Binned Phases}

For each sweep of the homodyne local oscillator phase, an accurate value of the output phase-shift $\Delta \phi_{DL}$ and the four-wave mixing phase-shift $\Delta \phi_{FWM}$ can be extracted from the peaks of the pulses (as illustrated in Figure 2b).
Therefore, for every output phase-shift (represented by a point in Figure 2c), we obtain an entire set of homodyne statistics $\{X(\theta)\}$, all associated with a particular FWM phase shift $\Delta \phi_{FWM}$. We then organize the output phase-shifts into 10 bins and then combine quadrature sets $\{X(\theta)\}$ with associated FWM phase-shifts in the same the bin region. A selection of these combined sets are plotted in Figure 3c.

\begin{figure}
\centering
\includegraphics[width=.9\linewidth]{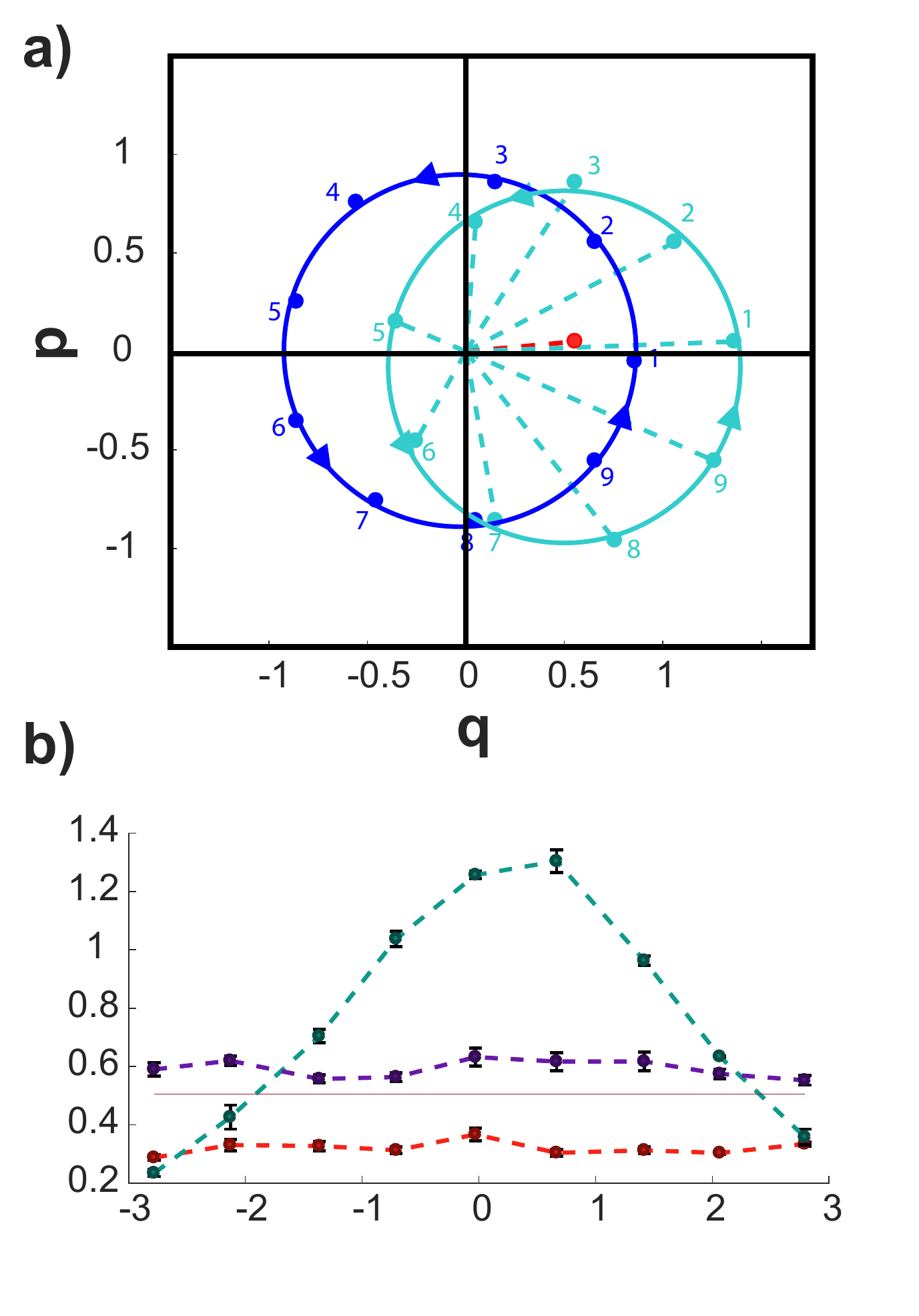}
\caption{ \textbf{a)} Maximum values of the Wigner functions of each post-selected phase are plotted, illustrating the motion in phase-space of the probe output as the global input phase is changed. These maximums are illustrated for the double-$\Lambda$ states (in teal), the four-wave mixing (in purple), and the probe-only case (in red) is labeled as a reference. \textbf{b)} Plots of mean photon number from reconstructed states, as a function of post-selected phase shifts. Photon numbers are plotted as dashed lines for double-$\Lambda$ (dashed teal), four-wave-mixing (dashed purple), EIT (dashed red), and probe input (solid red).}
\end{figure}

For each combined set (binned by $\Delta \phi_{FWM}$), we perform a quantum state reconstruction for each case: probe-only, double-$\Lambda$, and FWM. Additionally we also measure and reconstruct the density matrix for the probe and signal fields without the cell.

The reconstructed density matrices can additionally be mapped to a Wigner function representation for easier visualization. Figure 4 illustrates an input-output representation for different phase-shifted output states, while Figure 5a shows how both the four-wave mixing and the double-$\Lambda$ system traverse phase-space. In Figure 5a, we observe an unshifted circle (in dark blue) representing the max-values of each Wigner function of the four-wave mixing in phase space, indicating the mean photon number of the quantum state of the FWM does not change with phase. This is unlike the double-$\Lambda$ system which has its maximum values in phase space represented by a shifted circle (illustrated in teal), as expected by our semiclassical theory.

\subsection{High fidelity of output quantum state}
By comparing the quantum state of the phase-shifted output of the double-$\Lambda$, $\rho_{DL_{\theta}}$,  with the quantum state of the light without the cell, $\rho_{in_{\theta}}$, the fidelity can be calculated.
 \begin{equation}
F_{\theta}=\left(tr\sqrt{\sqrt{\rho_{in_{\theta}}}\rho_{DL_{\theta}}\sqrt{\rho_{in_{\theta}}}} \right)^2
\end{equation}

This input-output fidelity is plotted as a function of output phase shift (plotted as orange dashed lines in Figure 6a). This fidelity reaches its highest value of $94.4\%\pm .5\%$ for the bin $\Delta \phi_{FWM}\in (1.69,2.43$) radians.      The fidelity is also observed to reach $F =  91.9\% \pm 0.4\%$ for the bin $\Delta \phi_{FWM}\in (2.42,3.14$) radians.

Most of the behavior of the input-output fidelity can be explained with our semi-classical model. When the FWM is in-phase with the probe (ie, $\Delta\phi_{FWM} = 0$), the fidelity is lower because the quantum interference is constructive, producing a probe output is that larger than its input. When out-of-phase ($\Delta\phi_{FWM} = \pi$), the amplitude of the output flips in sign because of the unbalanced size of the probe and the four-wave-mixing generated by the signal field. These different cases can be seen in the homodyne sweeps illustrated in Figure 2b.

To be sure that our phase-shifted states do not obtain some additional quantum noise within our room-temperature system (such as Langevin noise \cite{PhysRevA.77.012323}), we measured the fidelity between our state and a theoretical coherent state of the same mean photon number (obtained from the reconstruction). As shown in Figure 6b, we find that our phase-shifted states retain remarkably high fidelity at the single-photon level - remaining above 90$\%$. Additionally, the purity of these quantum states were evaluated by both calculating the purity operator, $\rho^2$.  For the DL system exhibiting large phase-shifts (within the bin $\Delta \phi_{FWM} \in (2.42,3.14$) radians), we observe a purity of $0.92 \pm 0.02$, as compared to a purity of $0.93 \pm 0.01$ for the EIT-like case and $0.96 \pm 0.02$ for the FWM-only case. This high level of purity indicates that we are neither measuring significant parasitic thermal light nor decoherence in our phase-shift process.
\begin{figure}
\includegraphics[width=.9\linewidth]{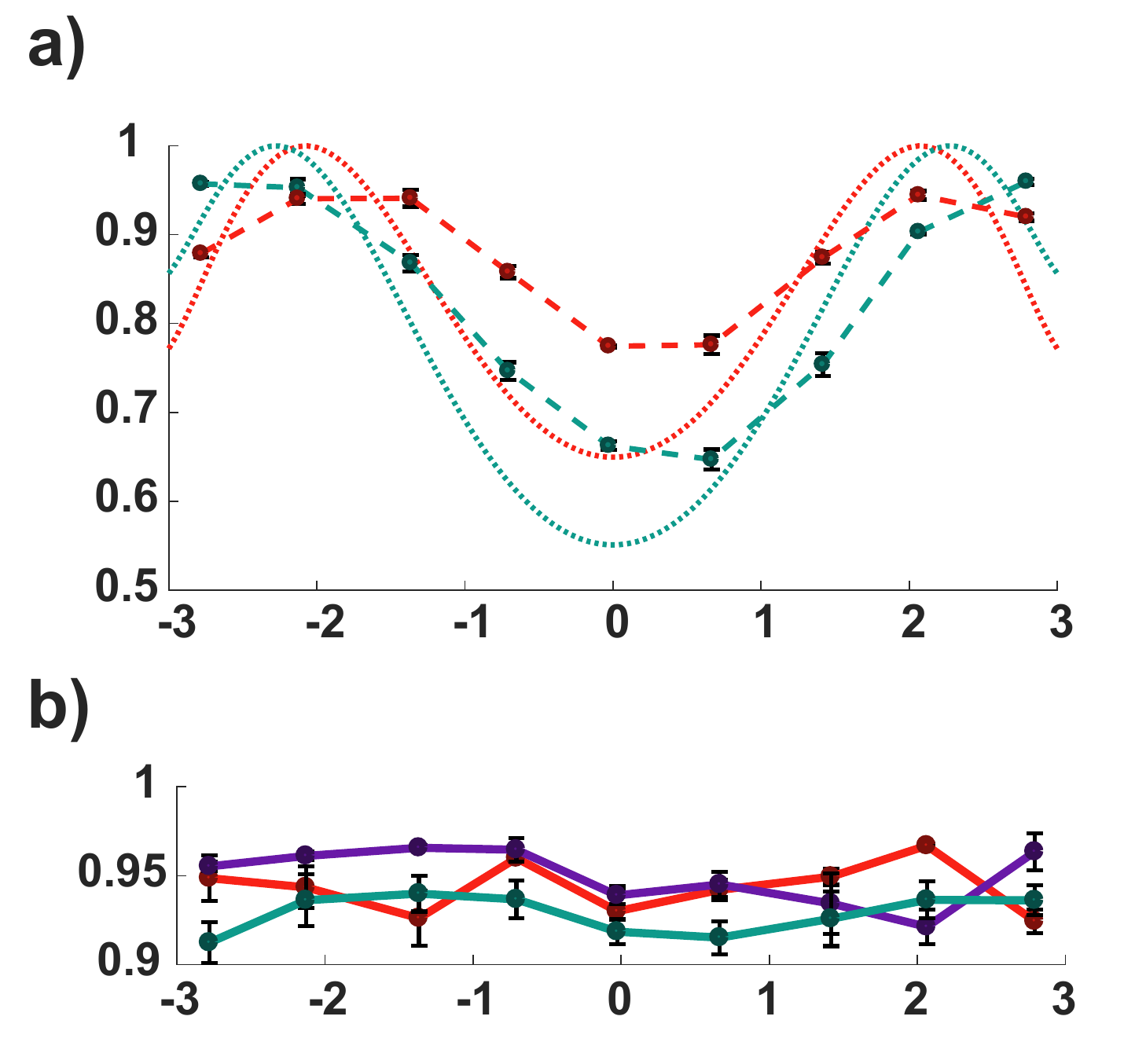}
\caption{\textbf{a)} Plots of fidelity from reconstructed states, as a function of post-selected phase shifts. The ``input-output" fidelity (plotted as a dashed red line) is found by comparing the density matrix overlap between a reference in which the probe is reconstructed without the Rb cell and the phase-shifted double-$\Lambda$ output state. The overlap between reconstructed states obtained between the pulse sequence 1 (probe-only case) and the pulse sequence 2 (double-$\Lambda$ case) is  plotted as a dashed green line. The expected fidelity plots for perfect coherent states are plotted for comparison (as dotted red and green lines).
\textbf{b)} Illustrating how similar the post-selected reconstructed states are to a perfect coherent state. For each post-selected phase-shift, the associated reconstructed state is compared to a theoretical, perfect coherent state of the same mean photon number. This is done for quadrature statistics for each pulse sequence (Probe-only, double-$\Lambda$ system, and FWM in red, teal, and purple respectively)}
\end{figure}

\section{Discussion}
In summary, we have shown that a simple room-temperature atomic system can create large phase-shifts in the quantum state of simultaneously propagating single-photon-level pulses; we have performed a quantum characterization of these phase-shifted output states using quantum state tomography; and we have demonstrated that four-wave mixing processes at room-temperature still have well-behaved quantum states with high fidelities, even for single-photon level light phase shifted by $\pi$. This bodes well for the future construction of FWM-mediated quantum nonlinear systems.

\subsection{Phase and Quantum State Estimation}

While a number of groups have characterized \cite{fwm_dressed} or implemented four-wave mixing \cite{PhysRevA.91.013843}, to our knowledge we are the first to reconstruct its Fock-state decomposition.
In particular, we note the utility of our pulse-sequence protocol, which enables a ``shot-by-shot" determination of the phase. While similar experiments require careful designs of the optical setup to reduce any potential change in path between the fields, our ``shot-by-shot" analysis allows us to accurately measure the phase with significantly less errors due to phase fluctuations (typically due to air fluctuations), thereby allowing for accurate quantum state estimation. This protocol could be useful for future experiments including: quasi-real-time characterzation of quantum light \cite{PhysRevLett.116.233602} and nondemolition measurements of single photons \cite{PhysRevLett.77.2352}.

Our results demonstrate that quantum protocols utilizing effective four-wave mixing processes in Rb cells are achievable, despite their room-temperature operation. The reconstructed states of the phase-shifted light resemble near-perfect coherent states. These states' fidelities (as compared to perfect coherent states) are as high as $94.4\% \pm 0.5 \%$ for the bin  $\Delta \phi_{FWM}\in (1.69,2.43)$ radians, a surprising result considering that thermal noise typically affects output fidelities in storage of light in similar systems. Additionally, for particular input-phases, we find the reconstructed quantum state of light's input-output fidelity is over 90$\%$.

\subsection{Truth-Table Elements for Gate Operation}

As discussed in the introduction, the traditional truth-table element of interest for gate operation is of the form:
   $$ (|0\rangle + |1\rangle)\otimes|1\rangle \Rightarrow (|0\rangle - |1\rangle)\otimes|1\rangle $$
where the triggering field is no longer a single-photon-level coherent state, but a single photon Fock state. While a complete measurement of this fidelity of this truth table element would require either Rubidium-tuned single photon sources or a dual-mode quantum process analysis \cite{Fedorov2015}, we believe our results allow us to answer many questions pertaining to the double-$\Lambda$ system's potential functionality for gate operation.

Our analysis of one of the two modes is equivalent to a reduction in the combined Hilbert space $H_A = H_P \otimes H_S$  \cite{Fedorov2015} to partial trace of the form: $\rho^P = Tr_S  \rho_{A}$. Although our probe mode has been reconstructed as a reduced density matrix $\rho^P$, because of the high level of purity of the reconstructed states, we can approximate our probe density matrix as a pure state. Therefore, for this case we can explicitly expand out the Fock-state elements of the probe mode after quantum process:
  \begin{align*}
      (|0\rangle + |\alpha_p| |1\rangle + ...) &\Rightarrow (|0\rangle + e^{i \Delta  \phi}|\alpha_p||1\rangle + ...) \\
      |\alpha_s\rangle &\Rightarrow |\tilde{\alpha_s}\rangle
  \end{align*}

where $\tilde{\alpha_p}$ and $\tilde{\alpha_s}$ represent coherent states that have modified phases and amplitudes for the probe and signal field respectively.
Extrapolating from Figure 5b, we approximate these values to be $|\alpha_p| = \sqrt{\langle N \rangle} \approx 0.71$ and $\Delta  \phi^p_{FWM} \approx1.86$ radians.

While these results appear to be immediately promising for quantum logic, some of the elements of the truth table necessary for quantum-phase gate operation are fundamentally limited. 
In this experiment, the primary mechanism for the phase-shift is due to phase-controllable frequency conversion via the FWM nonlinear process.  While quantum interference in our system is generated by a $\chi^3$ nonlinearity, the quantum coherence terms in $\rho_{12}$ generating four-wave mixing interfere linearly with respect to the input fields. This means that
  \begin{align}
  |0\rangle_p\otimes|1\rangle_s &\not\to |0\rangle_p\otimes|1\rangle_s \\
  &\to ((1-\epsilon)|0\rangle_p + \epsilon|1\rangle_p)\otimes|1\rangle_s
  \end{align}
The ``phase-triggering" photon $|1\rangle_s$ is partially converted into the other frequency mode, and conversion lowers the fidelity of the desired output state.

\section{Conclusion}
We conclude that while the current double-$\Lambda$ system cannot form a full ``truth table" necessary for quantum logic, the well-behaved nature of the analysed quantum states demonstrates that near-resonant atomic systems are a viable candidate for coherent-photon conversion even for bulk-glass cells operating at room-temperature.

Since the primary mechanism creating the phase-shifted light in this double-$\Lambda$ system are two interfering four-wave-mixing channels, it evaluates the first two benchmarks for implementation of the CPC protocol, as outlined in the introduction. For example, if parasitic nonlinear terms, spectral entanglement, or thermal effects dominate the process at the single photon level, then such a system will not be experimentally viable for a future CPC gate.

Observing no detrimental effects on fidelity of the quantum state of these 1-to-1 photon processes indicates that this architecture is ready to explore 1-to-2 photon conversion necessary for CPC \cite{Langford2011}. This would give our system the potential to achieve 2-qubit gate operations and quantum nondemolition measurements of single photons.

\section{Acknowledgements}
 The authors kindly thank Julio Gea Banacloche, Bing He, Balakrishnan Viswanathan, and Aephraim Steinberg for enlightening discussions. The work was supported by the US-Navy Office of Naval Research, grant number N00141410801, the National Science Foundation, grant numbers PHY-1404398 and PHY-1707919 and the Simons Foundation, grant number SBF241180. B. J. acknowledges financial assistance of the National Research Foundation (NRF) of South Africa. S. S. acknowledges financial assistance of the United States Department of Education through a GAANN Fellowship P200A150027-17.

\end{document}